\input{aipcheck}

\documentclass[
    ,final            % use final for the camera ready runs
  ]
  {aipproc}

\layoutstyle{8x11single}

\begin{document}

\title{Gamma-Ray Burst Groups Observed by Different Satellites}


\author{Istv\'an Horv\'ath}{
  address={Bolyai Military University, Budapest, Hungary}
}

\author{Lajos G. Bal\'azs}{
  address={Konkoly Observatory, Budapest, Hungary}
}

\author{P\'eter Veres}{
  address={Bolyai Military University, Budapest, Hungary}
  ,altaddress={Eotvos University, Budapest, Hungary} % additional visiting address
}

\begin{abstract}
 Two classes of gamma-ray bursts have been identified in the BATSE catalogs characterized by durations shorter and longer than about 2 seconds. There are, however, some indications for the existence of a third one. Swift satellite detectors have different spectral sensitivity than pre-Swift ones for gamma-ray bursts. Therefore it is worth to reanalyze the durations and their distribution and also the classification of GRBs. In this paper we are going to analyze the bursts' duration distribution and also the duration-hardness bivariate distribution, published in The First BAT Catalog, whether it contains two, three or maybe more groups. Similarly to the BATSE data, to explain the BAT GRBs duration distribution three components are needed. Although, the relative frequencies of the groups are different than they were in the BATSE GRB sample, the difference in the instrument spectral sensitivities can explain this bias in a natural way. This means theoretical models may have to explain three different type of gamma-ray bursts. 
\end{abstract}

\maketitle

\section{Introduction}

	It was always challenging to classify  gamma-ray bursts (GRBs).
 Using The First BATSE
Catalog, \citet{kou93} found a bimodality in the
distribution of the logarithms of the durations using $T_{90}$ (the time in which 90\% of the
fluence is accumulated \citep{kou93}) to characterize the
duration of GRBs. Today it is widely accepted that the physics of these two groups 
 are different, and these two kinds of GRBs are different phenomena \citep{nor01,bal03}.  The sky distribution of the
short bursts is anisotropic \citep{mbv00,vbh08}.
In the Swift database the two groups have a different redshift distribution, for short burst the
median is 0.4 \citep{os08} and for the long ones it is 2.4
\citep{bag06}.

Using the Third BATSE Catalog we have shown \cite{hor98} that the  duration ($T_{90}$) distribution
of GRBs observed by BATSE could be well fitted by a sum
of three log-normal distributions. 
We find it unlikely (with a probability $\sim 10^{-4}$) that there are only two
groups.  At the same time, \cite{muk98} report the
finding (in a multidimensional parameter space) of a very
similar group structure of GRBs.  Several authors
\citep{hak00,hak03,bor04,hak04,rkb05,chat07} included  more physical parameters into the analysis of the bursts (e.g. peak-fluxes, fluences, hardness ratios, etc.). 
A cluster analysis in this multidimensional parameter space suggests the existence of the third ("intermediate") group as well 
\citep{muk98,hak00,bala01,rm02,chat07}.  The physical existence of the third group is, however, still not convincingly proven. 
The celestial distribution of the third group is anisotropic \citep{mesz00,li01,mgc03}.   All these results  mean
that the existence of the third intermediate group in the BATSE
sample is acceptable, but its physical meaning, importance and
origin is less clear than those of the other groups. Thus, it is important
to investigate the presence of the third group in the Swift
sample as well.

\section{Duration and duration-hardness fits}

As in \citep{hor02} we fit the $\log T_{90}$ distribution using Maximum Likelihood (ML) method with a superposition of $k$ log-normal components. Each of the components have 2 unknown parameters to be fitted and the total number of measured points is $N=222$ (In the Swift First BAT Catalog there are 237 GRBs, of which 222 have duration information \citep{sak08}. For the distribution see Fig. 1. Our goal is to find the minimum value of $k$ suitable to fit the observed distribution. Assuming a weighted  superposition of $k$ log-normal distributions one has to maximize the following likelihood function:

\begin{equation}
L_k = \sum_{i=1}^{N} \log  \left(\sum_{l=1}^k   w_lf_l(x_i,\log
T_l,\sigma_l ) \right)
%\label{eq:ml}
\end{equation}

\noindent where $w_l$ are the weights, $f_l$ a log-normal function with
$\log T_l$ mean and $\sigma_l $ standard deviation having the form
of

\begin{equation}
f_l = \frac{1}{ \sigma_l  \sqrt{2 \pi  }}
\exp\left( - \frac{(x-\log T_l)^2}{2\sigma_l^2} \right)  
\label{fk}
\end{equation}

\begin{table}
\begin{tabular}{ccccc}
\hline
\tablehead{1}{c}{b}{centroid \\ in log(Duration)}
  & \tablehead{1}{c}{b}{$w$ (Swift BAT)}
  & 
  & \tablehead{1}{c}{b}{centroid in \\ log(Duration)}
  & \tablehead{1}{c}{b}{$w$ (CGRO BATSE)}   \\
\hline
-0.479 & 7.5\% & short & -0.301 & 20-26 \% \\
1.754 & 65\% & long & 1.565  & 60-66 \% \\
0.991 & 27.5\% & intermediate & 0.637  & 10-20 \% \\

\hline
\end{tabular}
\caption{The parameters of the three groups of GRBs}
\label{tab:a}
\end{table}

The three-Gaussian fit is shown in Figure 1. The best parameters were published in \citep{hor08}. 
Here in Table 1. we pesent the comparison of the parameters of the
GRB groups observed by CGRO BATSE and Swift BAT.
BAT
sensitivity is different than BATSE sensitivity was \citep{fis94,band03}.
BAT is more sensitive at low energies which means it can
observe more X-ray flashes and soft bursts and probably fails to detect many
hard bursts (typically short ones). Therefore one expects more
long and intermediate bursts and less short GRBs. In the BAT data
set there are only a few short bursts. Our analysis could only find
16 short bursts (7\%). The robustness of the ML method is demonstrated here
because a group with only 7\% weight is identified.
Previously, in the BATSE database the intermediate bursts were
identified by many research groups. However, in this class
different frequencies were found  representing 10-20 \% of BATSE GRBs
\citep{muk98,hak00,bala01,rm02,hor06}.

\begin{figure}
\caption{ Duration distribution of the First BAT Catalog 
gamma-ray bursts with a three-Gaussian fit.}
\includegraphics[width=0.6\textwidth]{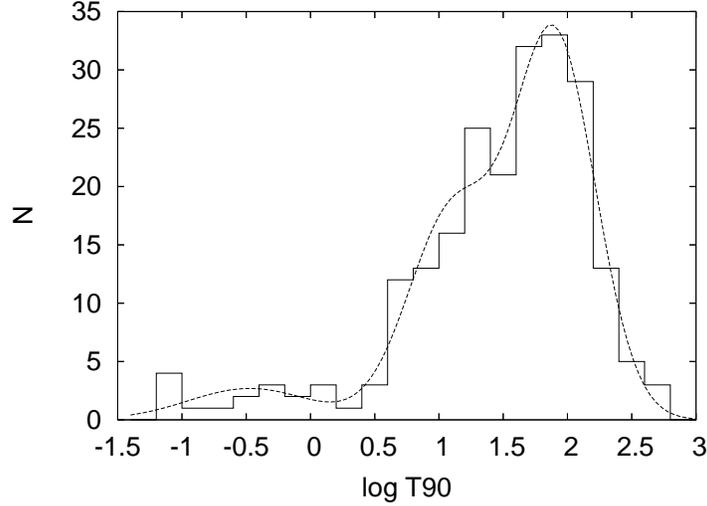}
\label{horvfig1}
\end{figure}

We can use more parameters for classification. The analysis of the clustering properties of GRBs in the BATSE 3B Catalog \cite{muk98} identified duration, total fluence and hardness as the relevant quantities for classification. Fitting the observed distribution with the superposition of Gaussian components one had to keep the number of estimated parameters as small as possible to ensure the stability of the Maximum Likelihood procedure, therefore we decided to use two dimensional Gaussians with the logarithmic duration and hardness. This work for the BATSE data was done by \cite{hor06} (Figure 2. left panel shows the result). 

Here we use the Swift First BAT catalog and fit 2D Gaussians on the duration hardness plane. Figure 2 right panel shows the result (black=short, light gray=intermediate and dark gray=long bursts). In Table 1. we compare the results of the Swift and BATSE data.

\begin{figure}
\caption{Classification in the duration-hardness plane (left BATSE, right Swift).}
\includegraphics[width=0.46\textwidth]{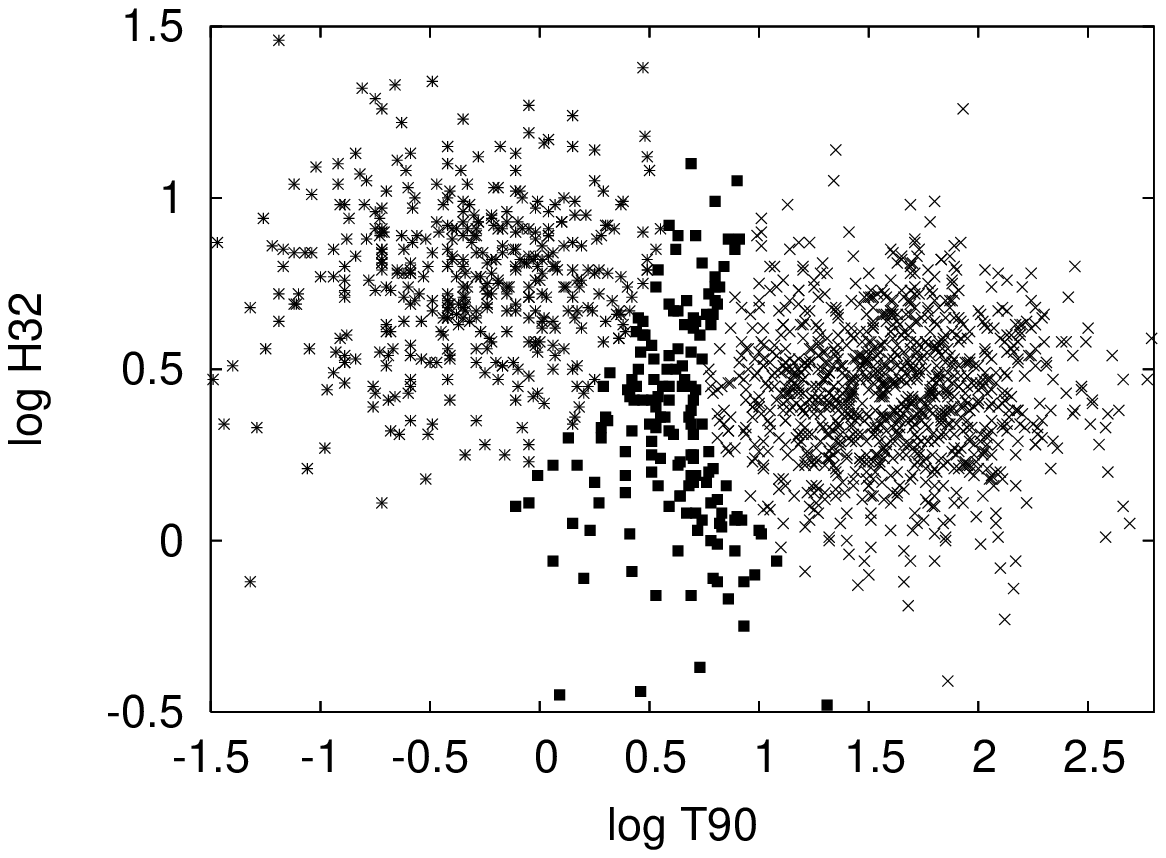}
\includegraphics[width=0.46\textwidth,height=0.25\textheight]{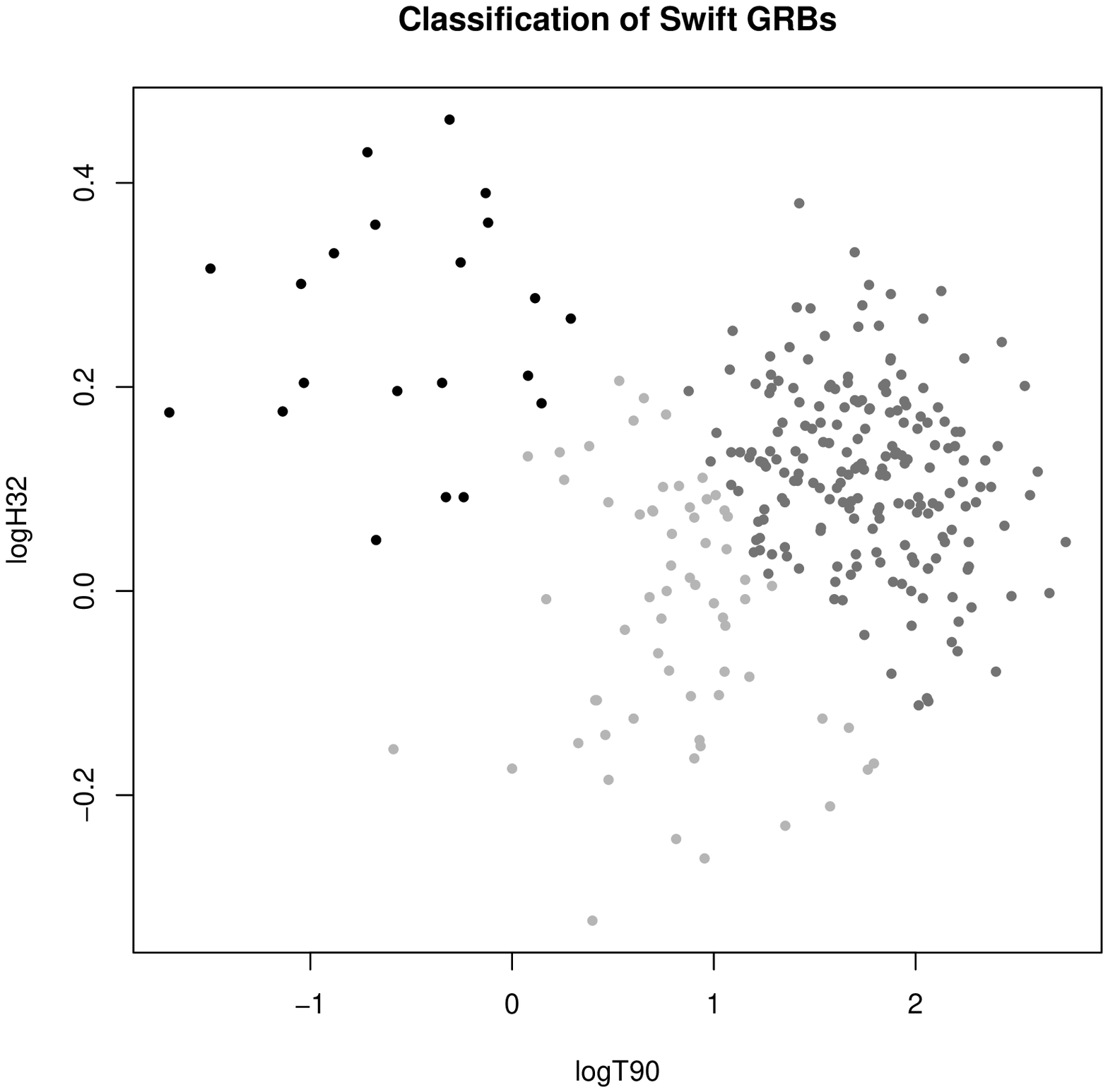}
\label{horvfig2}
\end{figure}

\begin{theacknowledgments}

  This research was supported in part through  OTKA T048870 and T077795 grants and Bolyai Scholarship (I.H.).

\end{theacknowledgments}

\bibliographystyle{aipproc}   % if natbib is available

\bibliography{horvath}

\end{document}